\documentclass[conference]{IEEEtran}
\ifCLASSINFOpdf
\else
\fi

\usepackage{balance}  % for  \balance command ON LAST PAGE  (only there!)
\usepackage{graphicx}
\usepackage{subfigure}
\usepackage{url}
\usepackage{mathtools}
\usepackage{underscore}
\usepackage{soul}
\usepackage{makeidx}
\usepackage{cite}
\usepackage{multirow}
\usepackage{colortbl}
\usepackage{epstopdf}

\begin{document}
\title{A Trust-based Recruitment Framework for Multi-hop Social Participatory Sensing}

\author{\IEEEauthorblockN{Haleh Amintoosi, Salil S.Kanhere}
\IEEEauthorblockA{School of Computer Science and Engineering\\
The University of New south Wales\\
Sydney, Australia\\
\{haleha,salilk\}@cse.unsw.edu.au}
}

\maketitle

\begin{abstract}
The idea of social participatory sensing provides a substrate to benefit from friendship relations in recruiting a critical mass of participants willing to attend in a sensing campaign. However, the selection of suitable participants who are trustable and provide high quality contributions is challenging. In this paper, we propose a recruitment framework for social participatory sensing. Our framework leverages multi-hop friendship relations to identify and select suitable and trustworthy participants among friends or friends of friends, and finds the most trustable paths to them. The framework also includes a suggestion component which provides a cluster of suggested friends along with the path to them, which can be further used for recruitment or friendship establishment. Simulation results demonstrate the efficacy of our proposed recruitment framework in terms of selecting a large number of well-suited participants and providing contributions with high overall trust, in comparison with one-hop recruitment architecture.
\end{abstract}


% \fi
\IEEEpeerreviewmaketitle

\section{Introduction}
\label{intro}
Great improvements in recent smartphone technologies, specially in terms of sensing capabilities, has resulted in the emergence of a new sensing paradigm, known as \emph{participatory sensing} \cite{Burke}. The main idea behind participatory sensing applications is involvement of ordinary citizens who collect sensor readings using their mobile phones. This revolutionary paradigm has been practically used to obtain information ranging from personal health \cite{DietSense, HealthSense} and prices of customer goods \cite{PetrolWatch} to environmental monitoring such as road conditions \cite{Hull} and noise pollution \cite{EarPhone}.

For a participatory sensing campaign to be a success, a number of challenges exist that should be addressed properly. The first challenge is the recruitment of sufficient participants, when there may be no explicit incentive. The second challenge is the suitability of participants to attend in the campaigns which require specific expertise or knowledge (such as taking photos of rare plant species) \cite{reddyrecruitment}. The third challenge is the trustworthiness of the contributions due to the open nature of participatory sensing \cite{Brian}.

To address these challenges in a comprehensive manner, one proposed idea is to leverage online social networks as the underlying infrastructure for participatory sensing applications \cite{integratemain}. The resulting paradigm, known as \emph{social participatory sensing} enables the usage of friendship relations for the identification and recruitment of participants.

In our previous work \cite{mobiquitous}, we addressed the challenge of assessing contribution trustworthiness by proposing a trust framework for social participatory sensing systems. In this framework, one-hop friends were selected as participants, and the trustworthiness of each contribution was calculated by considering the quality of contribution and trustworthiness of participant as two important trust aspects. These two aspects were then combined using fuzzy logic to reach to a trustworthiness score for each contribution, which was then used as a metric to select reliable contributions.

Leveraging immediate friends for participatory sensing data collection is beneficial since friends usually tend to be useful to their friends. However, in the absence of adequate friendship relations, the challenge of sufficient participants still exists. Consider a situation in which, a requester has few friends and not all of them attend in the sensing campaign. There might also be a case where a person has several friends but only a handful may contribute. In such cases, the validity of the derived information becomes a big concern due to the lack of sufficient contributions. In other words, with one-hop recruitment method, participant selection faces a potential constraint due to its limitation to the circle of immediate friends. This constraint will result in limited access to additional sources of data.

An extension to this recruitment method is to go deeper through the social graph and recruit friends of friends (FoFs) to attend in sensing campaigns. Such \emph{multi-hop} recruitment offers added values such as access to a large group of participants as valuable resources via friendship network of friends. It also increases the probability of access to well-suited participants with distinguished capabilities who are able to offer new perspectives and provide trustworthy contributions.

Developing a multi-hop recruitment strategy, however, opens up multiple questions. The first question being \emph{who should be recruited?} A trivial approach is to broadcast the task to all friends and FoFs. In this case, all friends and FOFs receive the task and are able to attend in the sensing campaign. However, in case of multi-hop recruitment, task flooding is not rational due to considerable overhead in terms of time, effort and network traffic. An alternative approach is to add a pre-selection process to the recruitment function, which is responsible for traversing through the requester's social graph (up to a specified level) and identifying suitable participants, i.e., those who can satisfy the task's requirements in an acceptable level. An indicator for this suitability is participant's performance in a past series of sensing campaigns. Participant's trustable contributions for several campaigns with specific knowledge or expertise requirements, demonstrates his suitability for future recruitment in similar tasks. The same stands for participants who always act in a timely manner, which makes them suitable candidates for attending in real-time contributions. Such valuable information should be extracted and used for recruitment of participants in future campaigns.

Another question in multi-hop recruitment is \emph{how participants should be recruited?} When suitable participants are selected, tasks should be sent from the requester to participants traversing social links. For a selected participant, there may be more than one route to reach. In this case, it is desirable to transfer the task from the most trustable path, to assure that the task's content and specifications has not been changed.

Since recruitment process is performed for each sensing campaign, one may wonder whether there is \emph{any suggested participant to be recruited in future campaigns?} In fact, sometimes a participant demonstrates an acceptable behaviour in terms of providing highly trustable contributions in a series of related campaigns. For example, he provides satisfiable contributions with high related expertise for a series of tasks which require taking photos of rare plant species. In such a case, it will be beneficial to add this participant to a suggestion list to be recruited for further campaigns.

In this paper, in order to fully address the above mentioned questions, we propose a trust-based recruitment framework for social participatory sensing systems. Our recruitment framework has four major components: The first component, called \emph{Participant Selection}, leverages multi-hop friendship relations in order to involve friends of friends in the sensing campaign and identify suitable candidates.  The second component, called \emph{Route Selection}, identifies the most trustable routes to selected participants. Once suitable participants are recruited, the third component, the \emph{Trust Server}, evaluates the contributions from selected participants and assigns a trust score that reflects the quality of the contributions. The forth component, called \emph{Suggestion Component}, provides the requester with a connection suggestion list containing a set of participants who have shown trustworthy behaviour in previous campaigns and are proper candidates for further recruitment or friendship establishment.

The rest of the paper is organised as follows. Related work is discussed in Section \ref{rel}. We present the details of our framework in Section \ref{pro}. Simulation results are discussed in Section \ref{sim}. Finally, Section \ref{con} concludes the paper.

\section{Related Work}
\label{rel}

\begin{figure*}
\centering
\includegraphics[height=5cm]{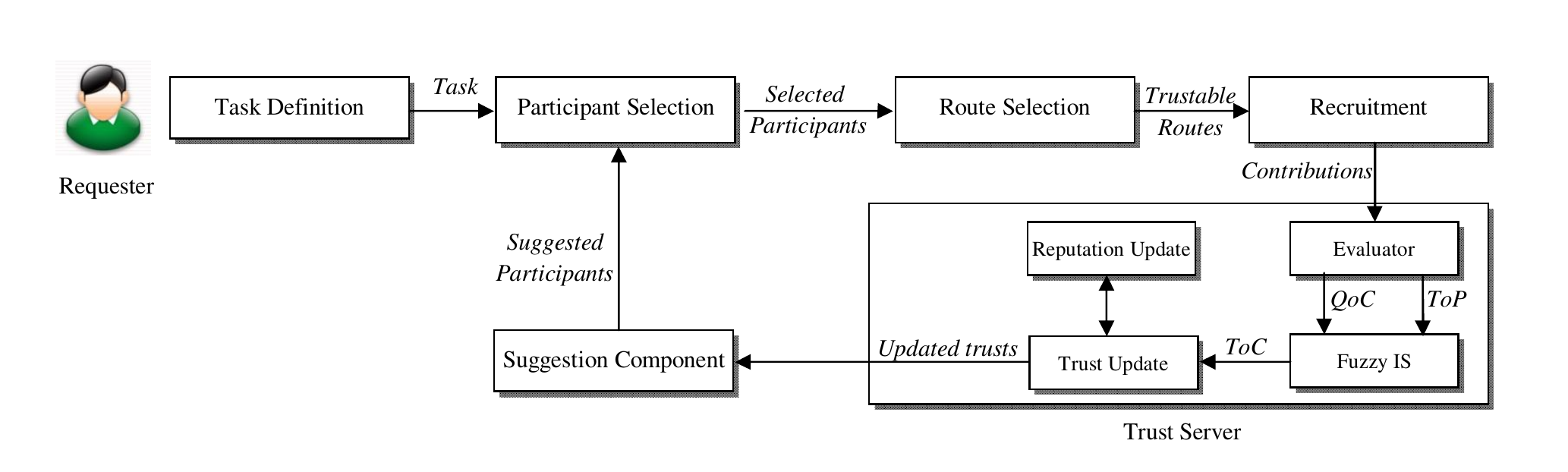}
\caption{Framework architecture}
\label{fig:frm}
\end{figure*}

To the best of our knowledge, the issue of participant recruitment in social participatory sensing hasn't been addressed in prior work. As such, we discuss about related research focussing on recruitment issues in participatory sensing and crowdsourcing. A typical recruitment framework receives the task requirements and provides the requester with a set of well-suited participants. Task requirements may involve a variety of factors including the task deadline, location of data collection, mobile device sensing capability and required expertise.

The process of identifying suitable participants is usually performed considering parameters such as availability, reputation and expertise, as described below.

\subsection{Availability-based Recruitment}

For most participatory sensing applications, availability of participants in terms of geographic and temporal coverage of the sensing area is important. For example, in a noise map sensing campaign, it is desirable to recruit participants who regularly pass through the sensing region and cover as much of the area as possible.
\cite{reddyrecruitment} proposed a recruitment framework for participatory sensing systems which aims at identifying suitable participants based on parameters such as geographical and temporal availability of participant. Participants have previously collected location traces for a period of time that represent their typical behaviour. The location traces are used by the qualifier component to select participants who have the task minimum location requirements. Once participants are selected, the assessment component identifies which subset of participants maximizes coverage over task specific area. These participants are then recruited and begin producing sensor data. During the campaign, the progress review component periodically monitors the coverage and availability of participants to see whether they remain consistent with task requirements.

\cite{distReq} also proposed a distributed recruitment and data collection framework for opportunistic sensing. The recruitment component exploits the suitability of user behaviours and based on the mobility history information, it recruits only the nodes that are likely to be in the sensing area when the sensing activity is taking place.  As a distributed recruitment framework, a set of recruiting nodes visit the sensor area before the campaign is launched and then disseminate recruitment messages. In order to transfer collected sensor data to the requester, a collection of nodes called data sinks are used and participating nodes opportunistically exploit ad hoc encounters to reach data sinks that are temporarily deployed in the sensed area.

\subsection{Reputation-based Recruitment}

Tracking participation in terms of performance has been employed widely by Internet businesses for commissioned work, such as Amazon Mechanical Turk \cite{Amazon} and GURU.com \cite{GURU}, which keep detailed statistics tracking the performance of requesters and workers. Moreover, auctioning systems such as e-Bay have transaction ratings to help evaluate whether a particular participant is trustworthy \cite{eBay}.

Narrowing it down to participatory sensing, authors of \cite{reddy} have presented a group of participatory sensing reputation metrics categorized into two groups: cross-campaign metrics such as number of previous campaigns taken and the success of participant in previous campaigns, and campaign-specific metrics such as timeliness, relevancy and the quality of sensor data. However, in \cite{reddyrecruitment}, they have limited reputation to considering participants' willingness (given the opportunity, is data collected) and diligence in collecting samples (timeliness, relevance and quality of data).

\cite{RFSN} proposes a reputation-based framework which makes use of Beta reputation \cite{beta} to assign a reputation score to each sensor node in a wireless sensor network. Beta reputation has simple updating rules as well as facilitates easy integration of ageing. However, it is less aggressive in penalizing users with poor quality contributions. A reputation framework for participatory sensing was proposed in \cite{Brian}. A watchdog module computes a cooperative rating for each device according to its short-term behaviour which acts as input to the reputation module which utilizes Gompertz function to build a long-term reputation score.

\subsection{Expertise-based Recruitment}

Expert based recruitment consists of identifying persons with relevant expertise or experience for a given topic. Expert finding have been excessively studied in social networks. \cite{expert1} developed a Bayesian hierarchical model for expert finding that accounts for both social relationships and content. The model assumes that social links are determined by expertise similarity between candidates. \cite{expert2} proposed a propagation based approach for finding expert in a social network. The approach consists of two steps. In the first step, they make use of person local information to estimate an initial expert score for each person and select the top ranked persons as candidates. The selected persons are used to construct a sub-graph. In the second step, one’s expert score is propagated to the persons with whom he has relationships.

Our recruitment framework is different from all above mentioned methods, since it considers social accountability of participants as an important factor in their reputation and selects well-suited participants who are highly reputable and satisfy the expertise and locality requirements of the tasks.

\section{Trust-based Recruitment Framework}
\label{pro}
In this section, we explain the proposed framework for trust-based recruitment in social participatory sensing systems. An overview of the architecture is presented in Section \ref{arc} followed by a detailed discussion of each component in Section \ref{comp}.

\subsection{Framework Architecture}
\label{arc}

Fig. \ref{fig:frm} illustrates the architecture of the proposed framework. The social network serves as the underlying publish-subscribe substrate for recruiting friends and FoFs as participants. In fact, the basic participatory sensing procedures (i.e., task distribution and uploading contributions) are performed by utilizing the social network communication primitives. We abstract an online social network (e.g., Facebook) as a weighted directed graph, in which, social network members serve as graph nodes and friendship relations denote the edges of graph, with weights equal to the trust rating between users. %Trust server is responsible for maintaining the social graph topology as well as relevant attributes for each user, derived from his public profile information.

A person wishing to start a participatory sensing campaign acts as a requester and defines the task, which includes the specification of task's main requirements such as needed expertise or location. Then, participant selection component utilizes the social graph to investigate requester's friends and FoFs as potential participants to see whether there is a match between the task requirements and participants' attributes. The result will be a set of participants who are suitable to attend in the sensing campaign.

Next, route selection component traverses the social graph to find the most trustable path from requester to each of selected participants. The task is then routed from the specified paths and delivered to the participants. The contributions received in response to a campaign are transferred (e.g., by using Facebook Graph API\footnote{http://developers.facebook.com/docs/reference/api/}) to a third party trust server which incorporates the fuzzy inference system and arrives at an objective trust rating for each contribution. Based on contributions' trust rating, mutual trusts along the route between participants and requester are updated.

In certain intervals (which is after attending in a certain number of campaigns), the suggestion component builds a custom contact group for each requester, which contains a list of participants who have shown a satisfiable performance in multiple campaigns. The list is further used for recruitment or friendship establishment. More details about each component is presented in section \ref{comp}.

\subsection{Framework Components}
\label{comp}
This section provides a detailed explanation of framework components. In particular, we go through the new components introduced in this paper, i.e., participant selection, route selection and suggestion component. We also present a short explanation of what we have done in our previous works \cite{mobiquitous, TechRep}.\\

\subsubsection{Participant Selection Component}

The participant selection component is responsible for detecting suitable participants. As said before, we assume that participants are social network members with public profile information as well as social links to their friends. In our previous work, we assumed that for each participant, two types of trust information are stored in the trust database: one is related to personal factors such as participant \emph{expertise}, his \emph{locality} score to different geographical areas and his \emph{timeliness} in responding to campaigns; the other one is related to social factors such as \emph{friendship duration} between participant and each of his friends, and the \emph{timegap} between their successive interactions.

When a task is defined, a set of requirements including the sensing region, deadline and required expertise is specified. In our previous work, we only assumed one-hop recruitment. So, the task was easily broadcast to all one-hop friends. Here, with the assumption of multi-hop recruitment, a pre-process step is necessary in order to find suitable participants who are able to satisfy the task's requirements. So, participant selection component performs a search in the related information of requester's friends and FoFs (up to a specified level) to detect suitable participants. Suitability is a general concept and depends on the requester's point of view. Here, we define suitability as an acceptable match between participant's trust factors and the task requirements. In particular, $P$ is regarded as a suitable participant if he has an acceptable (above a predefined threshold) locality score in the task region and acceptable expertise score in regards to the task's required expertise. These scores are evaluated and stored in the trust server (details in Section \ref{server}). The aforementioned thresholds are set by the requester while defining the task specifications and requirements.

The output of this component is a list of suitable participants who are eligible to be recruited in the campaign.\\

\subsubsection{Route Selection Component}

\begin{figure}
\centering
\includegraphics[height=4cm]{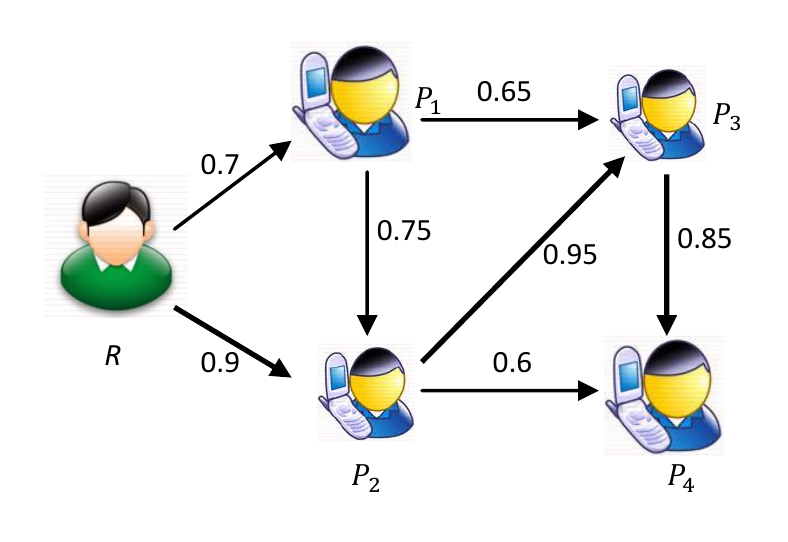}
\caption{a simple route selection example}
\label{fig:graph}
\end{figure}

Once suitable participants are selected, route selection component finds the most trustable routes from the requester to these participants. This process is quite similar to the problem of finding the best routes between two nodes in a directed graph. Consider the graph depicted in Fig. \ref{fig:graph}, in which, $R$ is the requester and $P_1$, $P_2$, $P_3$, and $P_4$ are participants; among which, $P_1$ and $P_4$ have been selected as suitable participants. As mentioned before, the edges of social graph are labeled with weights equal to the trust score between the nodes. For example, the weight of an edge from $R$ to $P_1$ is 0.7, showing the trust score of $R$ upon $P_1$. If there are intermediate nodes in the route from requester to a participant, trust score of the route is a combination of trust scores of each pair nodes along the route. We leverage multiplication as the combination function since it has been shown in \cite{multiplication} to be an effectiveness strategy for trust propagation. In this case, the trust score of the route from $R$ to $P_3$ (through $P_1$) will be 0.7*0.65=0.455.

Route selection component begins its search from the requester and tries to find the best route to each selected participant. We assume that the best route to immediate friends is the direct edge. For other participants, route selection component performs a depth first search through the social graph to reach to each selected participant. The search is finished when all routes from the requester to every suitable participant is defined. If more than one route is found between requester and a participant, the route with greatest trust score is selected as the most trustable route. For example, as shown in Fig. \ref{fig:graph}, there are 5 routes to reach $P_4$:\\
\\$r_1=R P_1 P_3 P_4$ \;    trust($r_1$)= $0.7*0.65*0.85 \simeq 0.39$\\
$r_2=R P_1 P_2 P_3 P_4$\; trust($r_2$)= $0.7*0.75*0.95*0.85 \simeq 0.42$\\
$r_3=R P_2 P_4$   \;      trust($r_3$)= $0.9*0.6=0.54$\\
$r_4=R P_2 P_3 P_4$   \;  trust($r_4$)= $0.9*0.95*0.85 \simeq 0.73$\\
$r_5=R P_1 P_2 P_4$   \;  trust($r_5$)= $0.7*0.75*0.6 \simeq 0.32$\\

The most trustable route between $R$ and $P_4$ is $r_4$.\\

\subsubsection{Trust Server}
\label{server}

When trustable routes to participants are defined, tasks are disseminated from requester to selected participants through selected routes. The trust server is responsible for maintaining and evaluating a comprehensive trust rating for each contribution, updating mutual trusts and calculating a reputation score for each participant. 

In our previous work \cite{mobiquitous,TechRep}, we assumed that there are two trust aspects that need to be considered: (1) Quality of Contribution (QoC) and (2) Trust of Participant (ToP). The server maintains a trust database, which contains the network topology as well as required information about participants and the history of their past contributions. When a contribution is received to the trust server, the effective parameters that contribute to the two aforementioned components are evaluated by the Evaluator and then combined to arrive at a single quantitative value in the range of [0, 1] for each. To evaluate QoC, we assumed of leveraging state-of-the-art methods such as image processing techniques (for image-based campaigns) and outlier detection techniques (for sound-based campaigns). As for ToP, personal parameters such as participant's \emph{expertise}, \emph{timeliness} and \emph{locality} as well as social parameters such as \emph{friendship duration} and \emph{interaction timegap} between requester and participant are evaluated and combined via a weighted sum aggregation function (The detailed description of each parameter along with its evaluation method has been thoroughly discussed in \cite{TechRep}). The two measures QoC and ToP serve as inputs for the fuzzy inference system, which computes the trustworthiness of contribution (ToC) in the range of [0, 1]. In fact, the fuzzy system is aimed at adjusting ToC according to different values of QoC and ToP.
 
It should be noted that ToP denotes the trustworthiness of participant from the requester's point of view. In case of one-hop recruitment, the requester is participant's immediate predecessor, which may not be always true in multi-hop recruitment. So, in order to correctly compute ToP in multi-hop recruitment, it should be combined with trustworthiness of the route from requester to the participant, before being fed to the fuzzy system. In other words, $ToP * trust(r)$ will be considered as ToP in which, $trust(r)$ is the trustworthiness score of the route from the requester to the participant.

Once ToC is calculated, contributions with ToC lower than a predefined threshold are revoked from further calculations. For certain campaigns, depending on the nature of task, the requester may desire to add a subjective evaluation in order to indicate how much the contribution is compatible with his needs and expectations. This subjective rating is referred to as Requester Evaluation (RE) and is in the range of [0, 1].

Since intermediate friends contribute in the task dissemination process, it is reasonable to update their trust scores at the end of each campaign. Consider a situation in which, there are $n-2$ intermediate friends $P_2$, $P_3$, $P_4$, ..., $P_{n-1}$  between $P_1$ and $P_n$. Similar to our previous work, we adopt a reward/penalty policy for this update. If the computed ToC value for participant $P_n$ is greater than a predefined threshold1$(Th_1)$, $P_n$ and all intermediate friends are rewarded, and the amount of $\left |ToC-\rho_{Req}*RE \right |$ is added to each of $Trust_{P_{n-1}P_n}$, in which, $\rho_{Req}$ is the reputation score of the requester. Similarly, if ToC value of $P_n$'s contribution is less than a predefined threshold2$(Th_2)$, he is penalized, and the amount of $\left |ToC-\rho_{Req}*RE \right |$ is reduced from $Trust_{P_{n-1}P_n}$. Trust update process can be summarized in Eq. \ref{eq:tr}, in which, $\gamma= \left |ToC-\rho_{Req}*RE \right |$. In our simulations in Section \ref{sim}, we set $(Th_1)=0.7$ and $(Th_2)=0.3$.

\begin{equation}
\label{eq:tr}
\footnotesize{Trust_{P_{i-1}P_i} =\left\{\begin{matrix}
Trust_{P_{i-1}P_i}+ \gamma & \; (\forall i=1..n) & \; if\, ToC>Th_1\\
Trust_{P_{i-1}P_i}- \gamma & \; (i=n) & \; if\, ToC<Th_2
\end{matrix}\right.}
\end{equation}

At regular intervals, a reputation score in the range of [0, 1] is also calculated for each participant, which is a combination of the trust ratings that requesters have assigned to him. This reputation score is further used as a weight for participant's evaluations, ratings or reviews. More details about the process of trust update and reputation score calculation can be found in \cite{TechRep}.\\

\subsubsection{Suggestion Component}

If $P_n$ demonstrates outstanding performance in multiple campaigns originated by $P_1$, it would be beneficial if a direct relation is established between them, since: \emph{i)} the time required for selecting $P_n$ as a suitable participant in further campaigns is reduced. \emph{ii)} less time and effort is consumed for task dissemination, since there is now only one-hop distance to $P_n$. \emph{iii)} an easier access to $P_n$'s friendship network is now available.

To provide $P_1$ with a suggestion list, the following process is performed by suggestion component:\\
For each participant $P_n$ who is not an immediate friend of $P_1$, a field called \emph{implicit trust} is kept. This field is initially set to zero and is updated whenever $P_n$ contributes to a task originated from $P_1$. The implicit trust update process is the same as trust update performed in trust server; i.e., it is increased by a constant amount $\left |ToC-\rho_{Req}*RE \right |$, if $P_n$ provides a contribution with ToC greater than threshold1, and decreased by the same amount if ToC is less than threshold2.

At certain intervals, implicit trust values are investigated to see whether $P_n$ is eligible to be suggested for recruitment or friendship establishment. If above a threshold, $P_n$ has such eligibility. In our simulation, we set this threshold to be 0.5.

In case of plenty of eligible participants, suggestion component chooses the best candidates among them. The best candidates are those participants who act as intermediate node in larger number of routes. Adding such candidates as immediate friends will cause a considerable reduction in path lengths to other participants.

The requester is then provided with a suggestion list which consists of participants' IDs, their implicit trusts, and the intermediate nodes along the way from the requester to participant. Requester utilizes this list for further recruitment or new friendship establishment with initial trust value equal to implicit trust.

\section{Experimental Evaluation}
\label{sim}
This section presents simulation-based evaluation of the proposed recruitment system. The simulation setup is outlined in Section \ref{setup} and the results are in Section \ref{res}.

\subsection{Simulation Setup}
\label{setup}
To undertake the preliminary evaluations outlined herein, we chose to conduct simulations, since real experiments in social participatory sensing are difficult to organise. Simulations afford a controlled environment where we can carefully vary certain parameters and observe the impact on the system performance. We developed a custom Java simulator for this purpose.

The data set that we use for our experiment is the real web of trust of Advogato.org \cite{Advogato}. Advogato.org is a web-based community of open source software developers in which, site members rate each other in terms of their trustworthiness. Trust values are one of the three choices master, journeyer and apprentice, with master being the highest level in that order. The result of these ratings among members is a rich web of trust, which comprises of 14,019 users and 47, 347 trust ratings. The distribution of trust values in the Advogato web of trust is as follows: master: 17,306, journeyer: 21,353, and apprentice: 8688. The instance of the Advogato web of trust referenced in this paper was retrieved on October 13, 2007. In order to conform the Advogato web of trust to our framework, we map the textual ratings in the range of [0, 1] as master = 0.8, journeyer = 0.6, and apprentice = 0.4.
%The number of vertices with no outgoing edges is 1, 152 and the number of vertices with no incoming edges is 738.

The Advogato web of trust may be viewed as a directed weighted graph, with users as the vertices and trust ratings as the directed weighted edges of the graph. So, it is in perfect match with our assumptions related to participants and their trust relations in social participatory sensing.

Whenever a task is launched, one of the Advogato users is selected to be the requester. Participant selection component traverses the Advogato graph beginning from the requester until L level to find suitable participants (L has been set to be 3).  Next, route selection component finds the most trustable paths to these participants. Tasks are then disseminated through selected routes and trust server calculates ToC for each receiving contribution. Trust ratings along the routes are then updated (details have been presented in Section \ref{comp}).

We run the simulation for 20 intervals, each consisting of launching 30 tasks. At the end of each interval, a list containing suggested friends is provided with each requester, who are recruited in further tasks. The list length is set to be 50.

We compare the performance of our framework against one-hop recruitment system described in our previous work \cite{mobiquitous}, which broadcasts the tasks to all immediate friends. We also consider a multi-hop recruitment system in which, no further friends are added. Specifically, we compare the following: (1) one-hop recruitment, which disseminates the task to all one-hop friends. (2) multi-hop recruitment without suggestion, in which, suitable participants are selected through multi-hop friendship relations and engaged in sensing campaign. No further friendship establishment is done. (3) multi-hop with FS: Our proposed framework, which is the same as the second one with suggestion component, which provides the requester with a set of suggested friends to be recruited in subsequent campaigns.

As mentioned in Section. \ref{comp}, a ToC rating is calculated for each contribution in trust server and those with ToC lower than a predefined threshold are revoked from further calculations. The ToCs for the non-revoked contributions are then combined to form an overall trust for that campaign. In other words, \begin{math} Overall Trust = \frac{\sum_{i=1}^{n} ToC}{n} \end{math} in which, \emph{n} is the number of non-revoked contributions. The revocation threshold is set to 0.5. We consider the overall trust as the evaluation metric. Greater overall trust demonstrates better ability to achieve highly trustable contributions and revoke untrusted ones. Overall trust has a value in the range of [0, 1].

Moreover, as we explained in Section \ref{comp}, the route selection component searches for the most trustable routes, which is the route with greatest trust score. This route, however, may not be the route with shortest length. Accessing trustable routes with shorter lengths is desirable since longer routes increase the task response time. Thus, in order to evaluate the performance of our recruitment framework in terms of finding the best routes, we compute Mean Route Trust (MRT). MRT is defined as sum of trust ratings of all routes from the requester to selected participants divided by sum of route lengths, where route length is denoted by the number of hops. Higher MRT demonstrates the better ability to find more reliable routes with shorter lengths. MRT is calculated after each campaign for all selected routes to all selected participants. MRT is a value in the range of [0, 1].

\subsection{Simulation Results}
\label{res}

\begin{figure}%[!h]
    \includegraphics[width=0.45\textwidth]{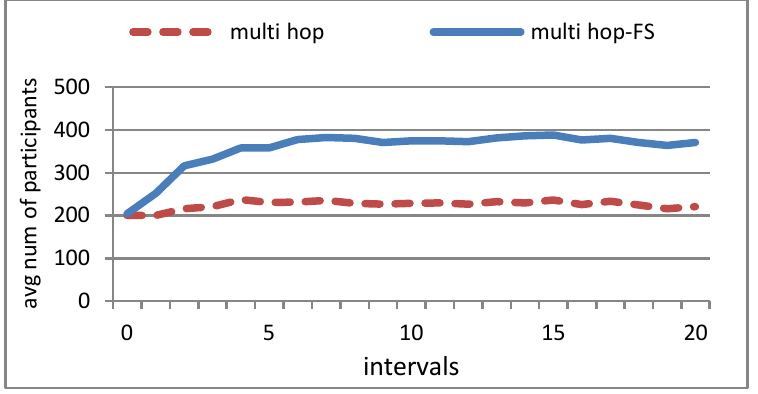}
    \caption{Evolution of average number of participants}
    \label{fig:nop}
\end{figure}

Figure \ref{fig:nop} demonstrates the average number of participants who have been selected and attended in the sensing campaigns inside each interval for both multi-hop and multi-hop with FS recruitment methods. As one-hop recruitment method has few participants in comparison with multi-hop, we omit its related data from this figure to better show the difference between two multi-hop recruitment methods. Note that interval 0 is in fact the first interval in which, no new friend has been suggested/added yet and hence, both multi-hop methods recruit equal number of participants.

As this figure shows, our proposed recruitment framework is able to recruit participants with an average number twice as multi hop recruitment method, which demonstrates a better performance in terms of finding and recruiting more suitable participants. Note that participant selection process in both multi-hop and multi-hop with FS recruitment methods is limited to L levels. But since our framework adds a set of suitable participants as immediate friends, an easier access to friends of these newly added friends (who may be located beyond the L levels) becomes available, which results in an increase in potential suitable participants. This demonstrates the effectiveness of our suggestion component in addressing the first challenge of participatory sensing systems described in Section \ref{intro}, which is recruiting adequate participants.
\begin{figure}%[!h]
    \includegraphics[width=0.45\textwidth]{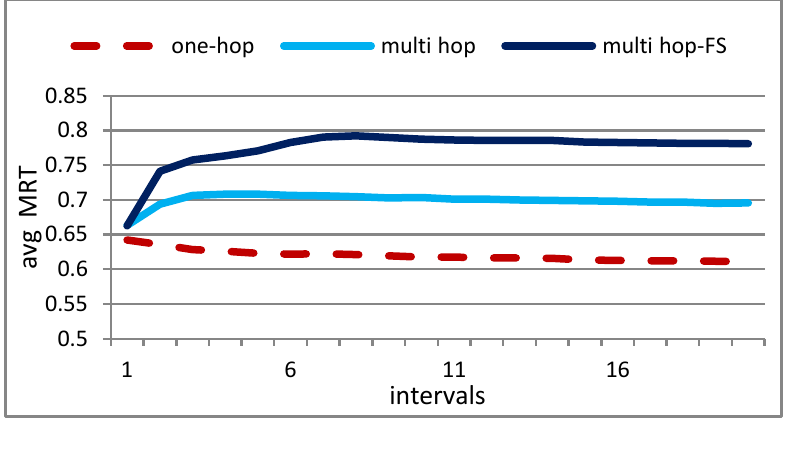}
    \caption{Evolution of average MRT}
    \label{fig:MRT}
\end{figure}

Figure \ref{fig:MRT} depicts the evolution of average MRT for 20 intervals. As this figure shows, the average MRT of our proposed recruitment framework is higher than the traditional multi-hop method. This means that our framework demonstrates better performance in finding trustable routes which, at the same time, have shorter lengths. One may argue that average MRT for one-hop recruitment method should be higher given that the route length in such method is less than route lengths in multi-hop recruitment methods. We would say that although route lengths in on-hop method are shorter, the sum of trust ratings for selected routes in multi-hop method is greater than one-hop method. The reason is that, as described in Section \ref{server}, \emph{all} the intermediate nodes along the route from the requester to participant are encountered an increase in their mutual trust ratings (in case of participant's satisfiable behaviour). This means that larger number of edges in multi-hop recruitment methods encounter such increase, as compared to one-hop method, which leads to higher MRT for multi-hop recruitment methods.

\begin{figure}%[!h]
    \includegraphics[width=0.45\textwidth]{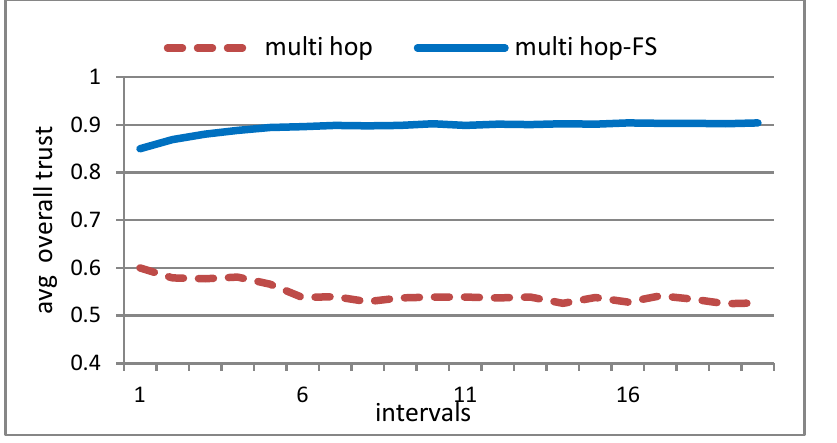}
    \caption{Evolution of average overall trust}
    \label{fig:overall}
\end{figure}

Finally, Figure \ref{fig:overall} demonstrates the evolution of average overall trust obtained from contributions with multi-hop recruitment methods. As this figure depicts, our proposed recruitment framework obtains higher average overall trust in comparison with other method. This is because our framework is able to identify more trustable routes to participants, which results in higher ToCs and hence, higher overall trusts. This clearly shows the success of our framework in recruiting suitable participant who produce trustworthy contributions.

To summarise, simulation results demonstrate a better performance for our multi-hop with FS framework in terms of achieving higher overall trust (0.4 greater than multi-hop method) with higher MRT.

\section{Conclusion}
\label{con}
In this paper, we proposed a trust based recruitment framework for social participatory sensing system. Our system leverages multi-hop friendship relations to identify well-suited participants. The system then provides trustable routes to selected participants by performing a depth first search in social graph. At specific intervals, requesters are supported with a set of suggested participants who can be recruited in further campaigns. Simulations demonstrated that our scheme increases the overall trust as compared to other methods, and provides trustable routes with shorter lengths for participant recruitment.


\begin{thebibliography}{1}
\balance

\bibitem{Burke}
J.~Burke, D.~Estrin, M.~Hansen, A.~Parker, N.~Ramanathan, S.~Reddy and M. B.~Srivastava, \emph{Participatory Sensing}, in WSW workshop, ACM SenSys'06, pp. 117-134, 2006.

\bibitem{DietSense}
S.~Reddy, A.~Parker, J.~Hyman, J.~Burke, D.~Estrin, and M.~Hansen, \emph{Image browsing, processing, and clustering for participatory sensing: lessons from a dietsense prototype}, in ACM EmNets'07, pp. 13-17, 2007.

\bibitem{HealthSense}
E. P.~Stuntebeck, J. S.~Davis II, G.D.~ Abowd, and M.~Blount, \emph{HealthSense: Classification of Health-related Sensor Data through User-assisted
Machine Learning}, in HotMobile'08, pp. 1–5, 2008.

\bibitem{PetrolWatch}
Y.~Dong, S.S. Kanhere, C.T. Chou, and N. Bulusu, \emph{Automatic Collection of Fuel Prices from a Network of Mobile Cameras}, in DCOSS'08, pp. 140-156. 2008.

\bibitem{Hull}
B.~Hull, V.~Bychkovsky, Y.~Zhang, K.~Chen, M.~Goraczko, A.~Miu, E.~Shih, H.~Balakrishnan, and S.~Madden, \emph{Cartel: a distributed mobile sensor computing system}, in ACM SenSys'06, pp. 125-138, 2006.

\bibitem{EarPhone}
R.K. Rana, C.T. Chou, S.S. Kanhere, N.~Bulusu, and W.~Hu, \emph{Ear-phone: an end-to-end participatory urban noise mapping}, in ACM/IEEE IPSN'10, pp. 105-116, 2010.

\bibitem{reddyrecruitment}
S.~Reddy, D.~Estrin, and M.~Srivastava, \emph{Recruitment Framework for Participatory Sensing Data Collections}, in Pervasive'10, pp. 138-155, 2010.

\bibitem{Brian}
K.L.~Huang, S.S.~Kanhere, and W.~Hu, \emph{On the need for a reputation system in mobile phone based sensing}, in Ad Hoc Networks, 2011.

\bibitem{integratemain}
I.~Krontiris and F.~Freiling, \emph{Urban Sensing through Social Networks: The Tension between Participation and Privacy}, in ITWDC'10, 2010.

\bibitem{mobiquitous}
H.~Amintoosi and S.S.~Kanhere, \emph{A Trust Framework for Social Participatory Sensing Systems}, in press, in MobiQuitous'12, 2012.

\bibitem{distReq}
G. S. Tuncay, G. Benincasa, and Ahmed Helmy, \emph{ Autonomous and distributed recruitment and data collection framework for opportunistic sensing},  in Mobicom'12, pp. 407-410, 2012.

\bibitem{Amazon}
“Amazon mechanical turk,” 2008. \url{http://www.mturk.com}

\bibitem{GURU}
“Guru.com - freelancers at online service marketplace.” 2008. \url{http://www.guru.com}

\bibitem{eBay}
 P.~Resnick, P.~Resnick, K.~Kuwabara, R.~Zeckhauser, and E.~Friedman, \emph{Reputation systems}, Communications of the ACM, vol. 43, no. 12, pp. 45–48, 2000.

\bibitem{reddy}
S.~Reddy, K.~Shilton, J.~Burke, D.~Estrin, M.~Hansen and M.~Srivastava, \emph{Evaluating participation and performance in participatory sensing}, in UrbanSense08, 2008.

\bibitem{RFSN}
S.~Ganeriwal, L.K.~Balzano, and M.B.~Srivastava, \emph{Reputation-based framework for high integrity sensor networks}, ACM TOSN'08, vol.~4, no. ~3, 2008.

\bibitem{beta}
B.E. Commerce, A.~Jøsang, and R.~Ismail, \emph{The beta reputation system}, in Bled Electronic Commerce Conference, 2002.

\bibitem{expert1}
Smirnova, Elena. \emph{A model for expert finding in social networks}, In Proceedings of the 34th international ACM SIGIR, pp. 1191-1192, 2011.

\bibitem{expert2}
J.~Zhang, J.~Tang, and J.~Li. \emph{Expert finding in a social network}, Advances in Databases: Concepts, Systems and Applications, Springer Berlin Heidelberg, pp. 1066-1069, 2007.

\bibitem{Advogato}
R.~Levien and A.~Aiken, \emph{Attack-resistant trust metrics for public key certification}, 7th USENIX Security Symposium. 1998.

\bibitem{TechRep}
Haleh Amintoosi, Salil S. Kanhere, \emph{Providing Trustworthy Contributions via a Reputation Framework in Social Participatory Sensing Systems}, Technical Report UNSW-CSE-TR-201304, University of New South Wales, January 2013. \url{ftp://ftp.cse.unsw.edu.au/pub/doc/papers/UNSW/201304.pdf}.

\bibitem{multiplication}
O.~Hasan, L.~Brunie, and J.~Pierson, \emph{Evaluation of the iterative multiplication strategy for trust propagation in pervasive environments}, in ACM ICPS'09, pp. 49-54, 2009.



\end{thebibliography}
\end{document}